\documentclass[11pt,a4paper,DIV=11,numbers=noenddot]{scrartcl}
\usepackage[utf8]{inputenc}
\makeatletter
\DeclareOldFontCommand{\rm}{\normalfont\rmfamily}{\mathrm}
\DeclareOldFontCommand{\sf}{\normalfont\sffamily}{\mathsf}
\DeclareOldFontCommand{\tt}{\normalfont\ttfamily}{\mathtt}
\DeclareOldFontCommand{\bf}{\normalfont\bfseries}{\mathbf}
\DeclareOldFontCommand{\it}{\normalfont\itshape}{\mathit}
\DeclareOldFontCommand{\sl}{\normalfont\slshape}{\@nomath\sl}
\DeclareOldFontCommand{\sc}{\normalfont\scshape}{\@nomath\sc}
\makeatother
\usepackage{epsf,amsmath,amssymb,graphicx,scalefnt,rotating,enumitem,fancyhdr}
\usepackage{array}
\usepackage{longtable}
\usepackage{slashed}
\usepackage{filemod}
\usepackage{multirow}
\usepackage[affil-it]{authblk}
\usepackage[usenames,dvipsnames]{color}
\usepackage[pdftex, colorlinks=true, linkcolor=black, filecolor=black, citecolor=black, urlcolor=black, draft=false, bookmarks, bookmarksnumbered=true, plainpages=false, linktocpage={true}]{hyperref}
\usepackage[final]{showlabels}

\usepackage[normalem]{ulem}
\usepackage[nospace,noadjust]{cite}
\newcounter{notecount}
\newcommand{\ringc}{\mathring{c}}
\newcommand{\three}{three}
\newcommand{\TF}{T_F}
\newcommand{\gcoup}[1]{\ifthenelse{\equal{#1}{}}%
                       {\frac{g}{4\pi}}{\frac{g^{#1}}{(4\pi)^{#1}}}}
\newcommand{\eom}{\abbrev{EOM}}
\newcommand{\emt}{\abbrev{EMT}}

\newcommand{\two}{two}
\newcommand{\one}{one}
\newcommand{\noeqn}[1]{(\ref{#1})}
\newcommand{\tcalo}{\tilde{\calo}}
\newcommand{\calo}{\mathcal{O}}
 
\newcommand{\citere}[1]{Ref.\,\cite{#1}}
\newcommand{\citeres}[1]{Refs.\,\cite{#1}}

\newcommand{\abbrev}[1]{{\scalefont{.9}#1}}
\newcommand{\EulerGamma}{\gamma_\text{E}}
\newcommand{\ep}{\epsilon}

\newcommand{\eqn}[1]{Eq.\,(\ref{#1})}
\newcommand{\eqs}[1]{Eqs.\,(\ref{#1})}
\newcommand{\fig}[1]{Fig.\,\ref{#1}}

\newcommand{\sct}[1]{Sect.\,\ref{#1}}

\newcommand{\dd}{\text{d}}
\newcommand{\deriv}[2]{\frac{\partial #1}{\partial #2}}
\newcommand{\dderiv}[2]{\frac{\dd #1}{\dd #2}}
\newcommand{\order}[1]{{\cal O}(#1)}

\newcommand{\qcd}{\abbrev{QCD}}

\newcommand{\lo}{\abbrev{LO}}
\newcommand{\nlo}{\abbrev{NLO}}
\newcommand{\nnlo}{\abbrev{NNLO}}

\newcommand{\msbar}{\ensuremath{\overline{\mbox{\abbrev{MS}}}}}

\newcommand{\RHheaderline}{TTK-18-32\\ May 2025%
}
\fancypagestyle{firstpage}
{
  
  \fancyhead[R]{\RHheaderline}
  \setlength{\headheight}{53pt}
}
\title{%
The two-loop energy-momentum tensor within the gradient-flow formalism}
\author{Robert V. Harlander}
\author{Yannick Kluth}
\author{Fabian Lange}
\affil{Institute for Theoretical Particle Physics and Cosmology,\protect\\
RWTH Aachen University, D-52056 Aachen, Germany}
\begin{document}
\date{}
\maketitle
\thispagestyle{firstpage}
\begin{abstract}
  The gradient-flow formulation of the energy-momentum tensor of
  \qcd\ is extended to \nnlo\ perturbation theory. This means that the
  Wilson coefficients which multiply the flowed operators in the
  corresponding expression for the regular energy-momentum tensor are
  calculated to this order. The result has been obtained by applying
  modern tools of regular perturbation theory, reducing the occurring
  two-loop integrals, which also include flow-time integrations, to a
  small set of master integrals which can be calculated analytically.
\end{abstract}
\newpage
\tableofcontents

\section{Introduction}\label{sec:intro}

The gradient-flow formalism as introduced by
L\"uscher~\cite{Luscher:2010iy} and further formalized by L\"uscher and
Weisz~\cite{Luscher:2011bx} has proven useful in lattice \qcd\ in many
respects. One of its main virtues is that composite operators at finite
flow time $t$ do not require ultra-violet (\abbrev{UV}) renormalization
beyond the one of the involved parameters and fields. This means that
the operators do not mix under renormalization-group running, which
makes it particularly simple to combine results from different
regularization schemes. This feature opens promising prospects for a
cross-fertilization of lattice and perturbative calculations, such as a
possible lattice determination of $\alpha_s(M_Z)$, for
example\,\cite{Harlander:2016vzb}.

A particularly powerful way to exhibit this possible interplay is
obtained by considering the expansion of composite operators in the
limit of small flow time, which expresses flowed operators in terms of
\qcd\ operators at $t=0$, with $t$-dependent Wilson
coefficients~\cite{Luscher:2011bx}. This method has been used by Makino
and Suzuki~\cite{Suzuki:2013gza,Makino:2014taa} to derive a
regularization-independent formula for the energy-momentum tensor (\emt)
$T_{\mu\nu}$ which has already led to promising results (see, e.g.,
\citeres{Asakawa:2013laa,Kitazawa:2016dsl,Taniguchi:2016ofw,Kitazawa:2017qab,
 Yanagihara:2018qqg}).

The universal Wilson coefficients that occur in the formula of
\citere{Makino:2014taa} for the \abbrev{EMT} have been calculated
through next-to-leading order (\nlo) in perturbation
theory~\cite{Suzuki:2013gza,Makino:2014taa}. This corresponds to a
\one-loop calculation in the sense that it involves integrals over a
single $D$-dimensional momentum. In this paper, we will carry this
calculation to the next perturbative order.\footnote{Note that we work
  in the limit of infinite volume. The inclusion of finite-volume
  effects requires different techniques, such as Numerical Stochastic
  Perturbation Theory, see \citere{DallaBrida:2017tru}.}  It is
important to note at this point that the integrals which occur in the
gradient-flow formalism are of a more general type than in regular
\qcd. They involve additional exponential factors which depend on loop
and external momenta, as well as on flow-time variables, some of which
are also integrated over. Nevertheless, the first two-loop result was
already obtained in \citere{Luscher:2010iy}, even in analytic form.  The
extension to the three-loop level required significant aid from computer
algebra and numerical tools\,\cite{Harlander:2016vzb}.  From the
quantum-field theoretical point of view, it closely followed the steps
of \citere{Luscher:2010iy} by directly expressing the Green's functions
in terms of integrals with the help of Wick's theorem. The integrals
themselves were evaluated using
sector-decomposition\,\cite{Binoth:2003ak,Smirnov:2013eza} in order to
isolate the poles in $D-4$, whose coefficients were determined using
high-precision numerical methods~\cite{genz,holodborodko}.

In the current calculation, we apply a completely independent setup.  On
the one hand, it applies the gradient-flow formalism described in terms
of a five-dimensional quantum field theory\,\cite{Luscher:2011bx}, which
leads to well-defined, albeit non-standard Feynman rules. On the other
hand, rather than evaluating the resulting integrals numerically, we
express them in terms of master integrals using the integration-by-parts
method of Chetyrkin and Tkachov\,\cite{Chetyrkin:1981qh}. This reduces
the \nlo\ calculation of the Wilson coefficients of the \emt\ to a
single \one-loop integral without flow-time integration. The
next-to-next-to-leading order (\nnlo) calculation leads to four
\two-loop master integrals without flow-time integration, and two
\two-loop master integrals with a single flow-time integration. All
master integrals can be calculated analytically by standard means for
general values of $D$, the number of space-time dimensions.

By suitable renormalization, the Wilson coefficients of the \emt\ can be
defined in such a way that they are formally renormalization-scale
independent. For a fixed-order perturbative result, this means that the
renormalization-scale dependence is formally of higher order. This
allows one to estimate the perturbative uncertainty on the Wilson
coefficients through their residual dependence on the renormalization
scale $\mu$ around a particular ``central'' value. Based on the form of
the analytical result, we argue for a specific choice of this central
value. Our numerical study shows that the higher-order terms indeed lead
to an appreciable reduction of the $\mu$-variation. However, by
comparison of the successive higher-order terms, it appears that the
uncertainty estimate from a variation within $\mu\in[\mu_0/2,2\mu_0]$,
as it is common practice in regular perturbative \qcd\ calculation,
might be too optimistic.

While we consider the \nnlo\ expressions for the Wilson coefficients of
the \emt\ as our main result, our calculation allows us to obtain a
number of additional results that might be useful in a broader
context. Among these are the flowed quark-field renormalization constant
$Z_\chi$, and the matrix of anomalous dimensions for the set of
operators which form the energy-momentum tensor in regular (non-flowed)
\qcd\ through \nnlo.

The remainder of the paper is structured as follows. After briefly
introducing the perturbative gradient-flow formalism in order to define
our notation in \sct{sec:pert}, we outline the approach of
\citeres{Suzuki:2013gza,Makino:2014taa} for using this formalism to
define the \emt\ in \sct{sec:emt}. Technical details of our calculation
are described in \sct{sec:calc}. Section\,\ref{sec:coefs} contains our
main result, the Wilson coefficients through \nnlo\ \qcd\ in the
\msbar\ scheme. As pointed out in \citere{Makino:2014taa}, the trace
anomaly of the \emt\ allows for a welcome check of the calculation; we
briefly describe the derivation of the resulting relations among the
coefficient functions in \sct{sec:traceanom}. Finally, in \sct{sec:zij}, we
use the finiteness condition of the flowed operators in order to derive
the anomalous-dimension matrix for the set of operators which occur in
the \emt\ in regular \qcd. Section\,\ref{sec:conclusions} presents our
conclusions.

\section{QCD gradient flow in perturbation theory}\label{sec:pert}

In the following, we will work in $D$-dimensional Euclidean space-time
with $D=4-2\ep$.  The gradient-flow formalism continues the gluon and
quark fields $A^a_\mu(x)$ and $\psi(x)$ of regular\footnote{We will use
  the terms ``flowed'' and ``regular'' \qcd\ to distinguish quantities
  defined at $t>0$ from those defined at $t=0$.} \qcd\ to
$(D+1)$-dimensional fields $B^a_\mu(t,x)$ and $\chi(t,x$) through the
boundary conditions
\begin{equation}
  \begin{split}
    B_\mu^a (t=0,x) = A_\mu^a (x)\,,\qquad \chi (t=0,x)= \psi(x)\,,
    \label{eq:bound}
  \end{split}
\end{equation}
and the flow equations
\begin{equation}
  \begin{split}
    \partial_t B_\mu^a &= \mathcal{D}^{ab}_\nu G_{\nu\mu}^b + \kappa
    \mathcal{D}^{ab}_\mu \partial_\nu B_\nu^b\,,\\ \partial_t \chi &= \Delta \chi
    - \kappa \partial_\mu B_\mu^a T^a \chi\,,\\ \partial_t \overline
    \chi &= \overline \chi \overleftarrow \Delta + \kappa \overline \chi
    \partial_\mu B_\mu^a T^a\,,
    \label{eq:flow}
  \end{split}
\end{equation}
where the ``flow time'' $t$ is a parameter of mass dimension minus two, and
$\kappa$ is an additional gauge parameter which drops out of physical
observables.

The $(D+1)$-dimensional field-strength tensor is defined as
\begin{align}
  G_{\mu\nu}^a = \partial_\mu B_\nu^a - \partial_\nu B_\mu^a + f^{abc}
  B_\mu^b B_\nu^c\,,
\end{align}
the covariant derivative in the adjoint representation is given by
\begin{align}
  \mathcal{D}_\mu^{ab} = \delta^{ab} \partial_\mu - f^{abc} B_\mu^c\,,
\end{align}
and
\begin{equation}
  \begin{split}
    \Delta = (\partial_\mu + B_\mu) (\partial_\mu + B_\mu)\,,\qquad
    \overleftarrow{\Delta} = (\overleftarrow\partial\!_\mu - B_\mu)
    (\overleftarrow\partial\!_\mu - B_\mu)\,.
  \end{split}
\end{equation}
As usual, the color indices of the adjoint representation are denoted by
$a,b,c,\ldots$, while $\mu,\nu,\rho,\ldots$ are $D$-dimensional Lorentz
indices. Color indices of the fundamental representation are suppressed
throughout this paper, unless required by clarity.  The symmetry
generators $T^a$ obey the commutation relation
\begin{equation}
  \begin{split}
    [T^a,T^b] = f^{abc}T^c\,,
  \end{split}
\end{equation}
with the structure constants $f^{abc}$.

The flow-field equation leads to a smearing of gauge-field
configurations at finite flow time $t>0$. As a consequence, composite
operators at $t>0$ do not require renormalization beyond the
renormalization of the parameters and fields of the Lagrangian. For the
strong coupling and the quark mass, the renormalization constants are
identical to those at $t=0$; the flowed gluon fields do not require
renormalization at finite flow time as was pointed out in \citere{Luscher:2011bx}.
The renormalization constant for the flowed quark field through
\nnlo\ is a by-product of this paper and will be given below.

\section{Energy-momentum tensor}\label{sec:emt}

In a continuous $D$-dimensional space-time, the gauge invariant part of the
\emt\ reads
\begin{equation}
  \begin{split}
    T_{\mu\nu}(x) \equiv \frac{1}{g_0^2}\left[
      \calo_{1,\mu\nu}(x)-\frac{1}{4}\calo_{2,\mu\nu}(x)\right]
    + \frac{1}{4} \calo_{3,\mu\nu}(x)
    - \frac{1}{2} \calo_{4,\mu\nu}(x)
    - \calo_{5 ,\mu\nu}(x) \,,
    \label{eq:emt}
  \end{split}
\end{equation}
where $g_0$ is the bare coupling constant of \qcd. The operators are
defined as
\begin{equation}
  \begin{split}
    \calo_{1,\mu\nu}(x) &\equiv
    F_{\mu\rho}^a(x)F_{\nu\rho}^a(x)\,,\\[10pt] \calo_{2,\mu\nu}(x)
    &\equiv
    \delta_{\mu\nu}F_{\rho\sigma}^a(x)F_{\rho\sigma}^a(x)\,,\\[10pt]
    \calo_{3_f,\mu\nu}(x) &\equiv
    \bar\psi_{f}(x)\left(\gamma_\mu\overleftrightarrow{D}\!_\nu
    +\gamma_\nu\overleftrightarrow{D}\!_\mu\right)\psi_{f}(x)\,,\\
    \calo_{4_f,\mu\nu}(x)
    &\equiv \delta_{\mu\nu}\bar\psi_{f}(x)
    \overleftrightarrow{\slashed{D}}\psi_{f}(x)\,,\phantom{\bigg(\bigg)}\\
    \calo_{5_f,\mu\nu}(x)
    &\equiv \delta_{\mu\nu}
    m_{f,0}\bar\psi_{f}(x)\psi_{f}(x)\,,\\
    \calo_{i,\mu\nu}(x) &=
    \sum_{f=1}^{n_F}\calo_{i_f,\mu\nu}(x)\,,\qquad i\in\{3,4,5\}\,,
    \label{eq:ops}
  \end{split}
\end{equation}
where $f$ labels the $n_F$ different quark flavors, $m_{f,0}$ is the
bare quark mass, and
\begin{equation}
  \begin{split}
    F^a_{\mu\nu} = \partial_\mu A_\nu^a - \partial_\nu A_\mu^a +
    f^{abc}A_\mu^bA_\nu^c\,,\qquad
    \overleftrightarrow{D}\!_\mu = \partial_\mu -
    \overleftarrow{\partial}\!_\mu + 2A_\mu\,.
  \end{split}
\end{equation}
The notation $i_f\in\{i_1,\ldots,i_{n_F}\}$ for the indices which label
different flavors will be useful later on in this paper.  In general,
$T_{\mu\nu}$ may contain gauge-dependent operators\, which vanish when
evaluating physical matrix elements \cite{Nielsen:1977sy}. Here and in
what follows, we implicitly assume that the vacuum expectations values
of all composite operators have been subtracted\footnote{In other words,
  the precise definition of $\calo_{1,\mu\nu}$, for example, would be
  given by $F_{\mu\rho}^aF_{\rho\nu}^a-\langle
  F_{\mu\rho}^aF_{\rho\nu}^a\rangle$.} so that $\langle
\calo_{i,\mu\nu}(x)\rangle\equiv 0\ \forall i$.

In this paper, we will focus on the case where physical matrix elements
of the \emt\ itself are considered, i.e.\ no other operator multiplies
the \emt\ at the same space-time point. In this case, the
equations-of-motion (\eom{}) render the set of operators in \eqn{eq:ops}
redundant. In particular, the \eom\ for the quark fields in regular
\qcd\ implies
\begin{equation}
  0 = \calo_{4, \mu \nu} (x) + 2 \, \calo_{5, \mu \nu} (x) \,,
  \label{eq:eom}
\end{equation}
which allows us to eliminate $\calo_{5,\mu\nu}$ from the set of
operators in \eqn{eq:ops}. Note that due to this relation, the last two
terms in \eqn{eq:emt} cancel.

We will further assume all quark masses to be equal to each other
\begin{equation}
	m_{f, 0} = m_0\,, \qquad f=1,\ldots,n_F\,.
\end{equation}
Therefore, the different quarks are indistinguishable and the
mixing between two different quark flavors cannot depend
on the flavors.

Defining the analogous operators of \eqn{eq:ops} for flowed fields, we
write
\begin{equation}
  \begin{split}
    \tcalo_{1,\mu\nu}(t,x) &\equiv
    G_{\mu\rho}^a(t,x)G_{\nu\rho}^a(t,x)\,,\\[10pt]
    \tcalo_{2,\mu\nu}(t,x) &\equiv
    \delta_{\mu\nu}G_{\rho\sigma}^a(t,x)G_{\rho\sigma}^a(t,x)\,,\\[10pt]
    \tcalo_{3_f,\mu\nu}(t,x) &\equiv Z_{\chi_f}
    \bar\chi_{f}(t,x)\left(\gamma_\mu\overleftrightarrow{\mathcal{D}}\!_\nu
    +\gamma_\nu\overleftrightarrow{\mathcal{D}}\!_\mu\right)\chi_{f}(t,x)\,,\\
    \tcalo_{4_f,\mu\nu}(t,x)
    &\equiv Z_{\chi_f} \delta_{\mu\nu}\bar\chi_{f}(t,x)
    \overleftrightarrow{\slashed{\mathcal{D}}}\chi_{f}(t,x)\,,
    \\
    \tcalo_{i,\mu\nu} &= \sum_{f=1}^{n_F}\tcalo_{i_f,\mu\nu}\,,\qquad i\in\{3,4\}\,,
    \label{eq:tops}
  \end{split}
\end{equation}
where $Z_{\chi_f}\equiv Z_\chi$ is the renormalization constant for the
flowed quark fields, and
\begin{equation}
  \begin{split}
    \overleftrightarrow{\mathcal{D}}\!_\mu = \partial_\mu -
    \overleftarrow{\partial}\!_\mu + 2B_\mu\,.
    \label{eq:leftrightcalD}
  \end{split}
\end{equation}
Since we have eliminated $\calo_{5,\mu\nu}$ from the set of operators by
using \eqn{eq:eom}, we do not need to include a flowed version of this
operator in \eqn{eq:tops}. Similar to the composite operators of regular
\qcd, we assume that the vacuum expectation values of the flowed
composite operators have been subtracted,
i.e.\ $\langle\tcalo_{i,\mu\nu}(t,x)\rangle\equiv 0\ \forall i$.

We can now use the expansion in small flow time\,\cite{Luscher:2011bx}
\begin{equation}
  \begin{split}
    \tcalo_{i,\mu\nu}(t,x) = \zeta_{ij}(t)\calo_{j,\mu\nu}(x) + \ldots\,,
    \label{eq:zetas}
  \end{split}
\end{equation}
to get a relation between flowed and regular \qcd\ operators. In
\eqn{eq:zetas}, and similarly in what follows, a sum $\sum_{j=1}^4$ is
understood. The ellipsis denotes terms that vanish as $t\to 0$ which
will be neglected throughout this paper. As discussed above, matrix
elements of the l.h.s.\ of this equation are finite after
renormalization of the \qcd\ parameters, while those of the regular
\qcd\ operators on the r.h.s.\ are in general divergent.  The mixing
matrix $\zeta_{ij}(t)$ will therefore be divergent as well.

Inverting \eqn{eq:zetas} and using it to re-express the regular
\qcd\ operators in the energy-momentum tensor in terms of flowed fields,
one arrives at
\begin{equation}
  \begin{split}
    T_{\mu\nu}(x) =  c_i (t) \tcalo_{i, \mu \nu} (t, x) \,,
    \label{eq:emtflow}
  \end{split}
\end{equation}
where
\begin{equation}
  \begin{split}
    c_i(t) \equiv \frac{1}{g_{0}^2} \left( \zeta_{1i}^{-1} (t)
			 - \frac{1}{4} \zeta_{2i}^{-1} (t) \right)
			 + \frac{1}{4} \zeta_{3i}^{-1} (t)
                         \,,\qquad i=1,\ldots,4\,.
    \label{eq:EMTGF}
	\end{split}
\end{equation}
Since matrix elements of the $\tcalo_i$ as well as the energy-momentum
tensor itself are finite (after mass and charge renormalization), the
universal coefficients $c_i(t)$ of \eqn{eq:EMTGF} are finite as
well.
In \citere{Makino:2014taa}, they have been calculated in perturbation
theory through \nlo\ \qcd. The goal of the current paper is to evaluate them
through \nnlo\ \qcd.

\section{Calculation of the Wilson coefficients}\label{sec:calc}

\subsection{Method of Projectors}

To compute the coefficients $\zeta_{ij}(t)$ we use the so-called
``method of projectors''\,\cite{Gorishnii:1983su,Gorishnii:1986gn},
which consists of constructing external states $|k\rangle$ and differential
operators $D_k$ for which
\begin{equation}
  \begin{split}
    P_{k}[\calo_{i} (x) ] \equiv D_{k}\langle
    0|\calo_{i} (x) |k\rangle =
    \delta_{ik}\,,
    \label{eq:projectors}
  \end{split}
\end{equation}
where we have dropped the Lorentz indices for convenience, and we define
the matrix element to include only diagrams which are one-particle
irreducible (1\abbrev{PI}) with respect to (w.r.t.)
\qcd\ particles. Applying $P_{k}$ on both sides of \eqn{eq:zetas}, one
obtains
\begin{equation}
  \begin{split}
    P_{k}[\tcalo_{i}(t,x)] =
    \zeta_{ij}(t)P_{k}[\calo_{j}(x)]\,.
    \label{eq:project}
  \end{split}
\end{equation}
Since the $\zeta_{ij}(t)$ only depend on the flow time $t$ and the
renormalization scale $\mu$, we can choose arbitrary values for all
other dimensional parameters in this equation. Setting them to zero
turns all higher-order corrections on the r.h.s.\ into massless
tadpoles, so that \eqn{eq:projectors} is only required to hold at
tree-level.  One thus obtains
\begin{equation}
  \begin{split}
    \zeta_{ij}(t) = P_j[\tcalo_i(t,x)]\Big|_{p=m=0}\,,
  \end{split}
\end{equation}
where $m$ and $p$ collectively denote all masses and external momenta.
The right-hand side thus results in vacuum diagrams whose only dimensional
scale is $t$.

In order to find suitable projectors, we first derive the Feynman rules
for the relevant terms of the operators. For example,\footnote{ All
  Feynman diagrams in this paper were drawn using Ti\textit{k}Z-Feynman
  \cite{Ellis:2016jkw}.}
\begin{align}
  \calo_{1,\mu\nu} &= \partial_\mu A_\rho^c\partial_\nu A_\rho^c +
  \cdots&\Rightarrow&&
  \begin{gathered}
	\includegraphics{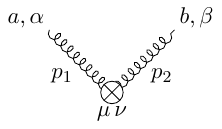}
  \end{gathered} =
  - p_{1,\mu} p_{2,\nu} \delta_{\alpha\beta}\delta^{ab},
\end{align}
where the momenta are defined to be outgoing.
This suggests to use
\begin{equation}
  \begin{split}
    P_{1}[X] = - \frac{\delta^{ab}}{N_A} P_{\alpha \beta |
      \rho \mu | \sigma \nu}
    \deriv{}{p_{1,\rho}}
    \deriv{}{p_{2,\sigma}}
    \langle 0 | A_\alpha^a (p_1) A_\beta^b (p_2) X_{\mu\nu}| 0 \rangle\,,
    \label{eq:p1}
  \end{split}
\end{equation}
where $N_A$ is the dimension of the adjoint representation of the gauge
group; for SU($N_c$), it is $N_A=N_c^2-1$. The projector onto the Lorentz structure is defined by
\begin{equation}
  \begin{split}
    P_{\alpha_1\beta_1|\cdots|\alpha_n\beta_n}
    T_{\alpha_1\beta_1\cdots\alpha_n\beta_n} = \left\{\begin{array}{ll}
    1\phantom{\Bigg(}\quad &\text{for}\quad
    T_{\alpha_1\beta_1\cdots\alpha_n\beta_n}=\delta_{\alpha_1\beta_1}
    \cdots\delta_{\alpha_n\beta_n}\,,\\[.3em]
    \multirow{2}{*}{0}& \text{for any other
      linearly}\\&\text{independent Lorentz tensor.}
    \end{array}\right.
  \end{split}
\end{equation}

In the appendix, one can find the relevant parts of the Feynman rules
for the other operators, which in a similar way lead to the projectors
\begin{equation}
  \begin{split}
    P_2[X] &=  - \frac{\delta^{ab}}{4 N_A} P_{\alpha \beta
      | \mu \nu | \rho \sigma}
    \deriv{}{p_{1,\rho}}
    \deriv{}{p_{2,\sigma}}
    \langle 0 | A_\alpha^a (p_1) A_\beta^b (p_2) X_{\mu \nu} | 0
    \rangle \,,\\
    P_{3_f}[X] &= -i \frac{\delta^{i j}}{4 N_c} P_{\rho
      \mu | \sigma \nu} \deriv{}{p_{2,\sigma}} \text{Tr} \left[
      \gamma_{\rho}
    \langle 0 | \psi_f^j (p_2) \bar{\psi}_f^i (p_1) X_{\mu \nu}
    | 0 \rangle \right]\,,\\
    P_{4_f}[X] &= -i \frac{\delta^{i j}}{4 N_c} P_{\mu \nu
      | \sigma \rho} \deriv{}{p_{2,\sigma}} \text{Tr} \left[
      \gamma_{\rho} \langle 0 | \psi_f^j (p_2) \bar{\psi}_f^i (p_1)
      X_{\mu \nu} | 0 \rangle \right] \\
    & \qquad - \frac{1}{2} \frac{\delta^{i j}}{4 N_c} P_{\mu
      \nu} \deriv{}{m_0} \text{Tr} \left[ \langle 0 | \psi_f^j (p_2)
      \bar{\psi}_f^i (p_1) X_{\mu \nu} | 0 \rangle \right]\,,
    \label{eq:pj}
  \end{split}
\end{equation}
where $N_c$ is the dimension of the fundamental representation of the
gauge group, i.e.\ the number of colors, and $i$ and $j$ are the corresponding indices.
The trace appearing in the projectors $P_{3_f}$ and
$P_{4_f}$ is taken w.r.t.\ the spinor indices of the Green's function,
and $f$ denotes the associated quark flavor.
Note that $P_{4_f}$ is constructed such that
\begin{equation}
  \begin{split}
    P_{4_f}[\calo_4 + 2\calo_5] = 0
  \end{split}
\end{equation}
in order to ensure that only those fermionic operators are taken into
account which do not vanish according to the \eom, see \eqn{eq:eom}.

With this procedure, we get a mixing matrix which distinguishes between
different quark flavors. To avoid confusion with $\zeta_{ij} (t)$, which
is the mixing matrix between operators summed over all flavors, we will
call it $\Omega_{ij} (t)$. This matrix is then defined by
\begin{equation}
	\tcalo_{i, \mu \nu} (t, x) = \Omega_{i j} (t) \calo_{j, \mu \nu} (x) \, ,
\end{equation}
where double indices in this expression are summed over $\{1, 2,
3_1,\ldots, 3_{n_F}, 4_1,\ldots,4_{n_F}\}$ (see \eqn{eq:ops}), in
contrast to \eqn{eq:emtflow}, where double indices are summed over $\{1,
2, 3, 4\}$. Its general structure is given by
\begin{equation}
	\Omega =
	\begin{pmatrix}
		\Omega_{1 1} & \Omega_{1 2} & \underline{\Omega_{1 3}}^T & \underline{\Omega_{1 4}}^T \\
		\Omega_{2 1} & \Omega_{2 2} & \underline{\Omega_{2 3}}^T & \underline{\Omega_{2 4}}^T \\
		\underline{\Omega_{3 1}} & \underline{\Omega_{3 2}} & \underline{\underline{\Omega_{3 3}}} & \underline{\underline{\Omega_{3 4}}} \\
		\underline{\Omega_{4 1}} & \underline{\Omega_{4 2}} & \underline{\underline{\Omega_{4 3}}} & \underline{\underline{\Omega_{4 4}}}
	\end{pmatrix} \, ,
	\label{eq:Omega}
\end{equation}
where an element $\Omega_{i j}$ represents the mixing between $\calo_i$
and $\tcalo_j$, taking into account individual flavors. Therefore, an
underlined and a double underlined element denotes an $n_F$-dimensional
vector and an $n_F \times n_F$ dimensional matrix, respectively.  Their
elements describe the mixing between different flavors.  As the quarks
are indistinguishable, the $n_F$ and $n_F^2$ dimensional objects
appearing in \eqn{eq:Omega} can each be described by two independent
parameters, named $\omega_{ij}$ and $\bar{\omega}_{ij}$:
\begin{equation*}
	\begin{split}
		\Omega_{i j} = \omega_{i j} \quad \text{for} \quad i,j <
                3, \qquad
		\underline{\Omega_{i j}} = \omega_{i j}
		\begin{pmatrix}
			1 \\
			\vdots \\
			1
		\end{pmatrix} \quad \text{for} \quad i < 3,\
                j > 2 \quad \text{or} \quad i > 2,\ j < 3\,,
	\end{split}
\end{equation*}
\begin{equation}
	\begin{split}
		\underline{\underline{\Omega_{i j}}} =
		\begin{pmatrix}
			\omega_{i j} & \overline{\omega}_{i j} &
                        \overline{\omega}_{i j} & \dots &
                        \overline{\omega}_{i j} \\
			\overline{\omega}_{i j} & \omega_{i j} &
                        \overline{\omega}_{i j} & \dots &
                        \overline{\omega}_{i j} \\
			\overline{\omega}_{i j} & \overline{\omega}_{i j}
                        & \omega_{i j} & \dots & \overline{\omega}_{i j}
                        \\
			\vdots & \vdots & \vdots & \ddots & \vdots \\
			\overline{\omega}_{i j} & \overline{\omega}_{i
                          j} & \overline{\omega}_{i j} & \dots &
                        \omega_{i j}
		\end{pmatrix} \quad \text{for} \quad i,j > 2\, .
	\end{split}
\end{equation}
Summing over the different flavors occurring in \eqn{eq:Omega}, the
relation between $\Omega (t)$ and $\zeta (t)$ can be easily established
by
\begin{align}
	\zeta_{i j} &= \omega_{i j} & \text{for} &\quad i < 3 \, , \\
	\zeta_{i j} &= n_F \, \omega_{i j} & \text{for} &\quad i > 2, \, j < 3 \, , \\
	\zeta_{i j} &= \omega_{i j} + (n_F - 1) \, \overline{\omega}_{i j} & \text{for} &\quad i > 2, \, j > 2 \, .
\end{align}

\subsection{Computational Methods}

The gradient-flow formalism in perturbation theory can be formulated in
terms of a Lagrangian field theory, where the flow equations
(\ref{eq:flow}) are implemented with the help of Lagrange-multiplier
fields \cite{Luscher:2011bx}. The crucial difference between the regular
\qcd\ Feynman rules and those in the gradient-flow formalism is the
occurrence of exponential factors $\exp(-sp^2)$, where $s$ is a
``flow-time variable'', and $p$ the linear combination of
$D$-dimensional external and/or loop momenta. Vertices involving flowed
fields induce an integration over all positive values of the
corresponding flow-time variable, which is, however, bounded from above
by ``propagators'' of the Lagrange-multiplier fields, since they
introduce step functions of the flow-time variables.

We have implemented the Feynman rules into the program
\texttt{qgraf}\,\cite{Nogueira:1991ex,Nogueira:2006pq}, which generates
the Feynman diagrams for the desired matrix elements. Its output is then
transformed to \texttt{FORM}\,\cite{Vermaseren:2000nd,Kuipers:2012rf}
notation by
\texttt{q2e/exp}\,\cite{Harlander:1997zb,Seidensticker:1999bb}. An
in-house set of \texttt{FORM} routines inserts the Feynman rules,
performs the projections onto the relevant color and Lorentz structures
according to the $P_j$ of \eqn{eq:p1} and \noeqn{eq:pj}, and evaluates
the Dirac and color traces using the \texttt{color}
package\,\cite{vanRitbergen:1998pn}.  The result is then expressed in
terms of a linear combination of integrals whose general form is as
follows:
\begin{equation}
  \begin{split}
    I_l( &(d_1, \dots, d_f ), (b_1, \dots, b_n ), ( a_1, \dots, a_n ))
    \\ &\equiv \frac{1}{\pi^{lD/2}} \, t^{lD/2-\sum_{j=1}^n a_j} \left[\prod_{i=0}^f
     \int_0^1 \mathrm{d}u_i u_i^{d_i} \right]
    \left[\prod_{r=1}^l\int\dd^Dk_r\right]\frac{\exp(-t \sum_{j=1}^n b_j
      q_j^2)}{(q_1^2)^{a_1}\dots (q_n^2)^{a_n}}\,,
    \label{eq:fl}
  \end{split}
\end{equation}
where the $a_i$ and $d_i$ are integers ($d_i\geq 0$), $f$ and $l$ is the
number of flow-time and loop integrations, respectively, the $b_j$ are
polynomials in (rescaled) flow-time parameters $u_i$ and the $q_i$ are
linear combinations of the loop momenta $k_j$. For the problem and the
perturbative order under consideration, it is $0\leq f\leq 4$, $1 \leq
l\leq 2$, and $0 \leq n \leq 3$, respectively.  Note that the projectors
defined in \eqs{eq:p1} and \noeqn{eq:pj} eliminate all dependence on
external momenta and masses, so that, after making a suitable ansatz for
the index structure of the integrals, we only have to evaluate scalar
vacuum integrals.  Using the identities\,\cite{Chetyrkin:1981qh}
\begin{equation}
  \begin{split}
    &\int\dd^D k \left(\deriv{}{k}\cdot q\right) f(k,q,\ldots) =
    0\,,
    \label{eq:ibp}
  \end{split}
\end{equation}
and similar ones for the flow-time integrations,
\begin{equation}
  \begin{split}
    \qquad\int_0^1\dd s \deriv{}{s} f(s,\ldots) =
    f(1,\ldots)-f(0,\ldots)\,,
    \label{eq:ibpflow}
  \end{split}
\end{equation}
one can derive relations among these integrals by explicitly performing
the derivatives in the integrand on the l.h.s. These so-called
``integration-by-parts (\abbrev{IBP}) relations'' were fed to
\texttt{Kira}\,\cite{Maierhoefer:2017hyi} which allowed us to reduce all
occurring integrals to a single master integral at \one-loop level, and
six master integrals at \two-loop level using the Laporta
algorithm\,\cite{Laporta:2001dd}.\footnote{The reduction with
  \texttt{Kira} 1.0 takes about 20 minutes on 8 CPU threads and requires
  less than 13\,GB of RAM.}  Their analytical evaluation is possible
along the lines of \citere{Luscher:2010iy}:
\begin{equation}
  \begin{split}
    I_1((),(2),(0)) &= 2^{-D/2} \,,\phantom{\Bigg[} \\[10pt]
    I_2((),(0,2,2),(0,0,0)) &= 2^{-D} \,,\phantom{\Bigg[} \\[10pt]
    I_2((),(1,1,1),(0,0,0)) &= 3^{-D/2} \,,\phantom{\Bigg[} \\[10pt]
    I_2((),(1,1,1),(1,1,0)) &= \frac{1}{D-2} \Bigg[ - 2 \pi \csc \left( \frac{D \pi}{2} \right) \\
    & \qquad \qquad+ \frac{3^{2 - D/2}}{D - 4} \, {_2F}_1 \left(1, 1; 3 - \frac{D}{2}; \frac{3}{4} \right) \Bigg] \,,\\[10pt]
    I_2((),(0,0,2),(1,1,0)) &= - \frac{2^{3 - D} \pi}{D - 2} \csc \left( \frac{D \pi}{2} \right) \,, \phantom{\Bigg[} \\[10pt]
    I_2((0),(2-u_1,u_1,u_1),(0,0,0)) &= 2^{2 - 2D} B_{1/4} \left( 1 - \frac{D}{2}, 1 - \frac{D}{2} \right) \,,\phantom{\Bigg[} \\[10pt]
    I_2((0),(1+u_1,1+u_1,1-u_1),(0,0,0)) &= 2^{2 - 2 D} \Bigg[ B_{3/4} \left( 1 - \frac{D}{2}, 1 - \frac{D}{2} \right) \\
    & \qquad \qquad - B_{1/2} \left( 1 - \frac{D}{2}, 1 - \frac{D}{2} \right) \Bigg] \,.
    \label{eq:masters}
  \end{split}
\end{equation}
In these expressions, we used $\csc(z)=1/\sin(z)$, and the
hypergeometric function defined as
\begin{equation}
  \begin{split}
    _2F_1(a,b;c;z) \equiv
    \sum_{n=0}^\infty\frac{(a)_n(b)_n}{(c)_n}\frac{z^n}{n!}\,,
  \end{split}
\end{equation}
with the Pochhammer symbol
\begin{equation}
	(x)_n \equiv \frac{\Gamma ( x + n)}{\Gamma(x)} \,.
\end{equation}
Furthermore, the incomplete beta function is defined by
\begin{equation}
  B_z (a, b) \equiv \int_0^z\dd t\, t^{a-1}(1-t)^{b-1}
\end{equation}
and can be expressed as
\begin{equation}
  \begin{split}
    B_z(a,b) =  z^a \sum_{n = 0}^\infty \frac{(1 - b)_n}{n! (a + n)} z^n
    = \frac{z^a}{a}\ {}_2F_1(a,1-b;a+1;z)\,.
  \end{split}
\end{equation}
The expansions of the hypergeometric function in the limit $\ep\to 0$
can be obtained with the help of the
\texttt{Mathematica}\,\cite{Mathematica11.3} package
\texttt{HypExp}\,\cite{Huber:2005yg,Huber:2007dx}.

A more detailed description of parts of our setup will be described in a
forthcoming publication\,\cite{Artz:2019bpr}. As a check, we evaluated the
correlators $\langle G_{\mu\nu}^aG_{\mu\nu}^a\rangle$, $\langle
\bar\chi\chi\rangle$ and $\langle \bar\chi \slashed D \chi\rangle$
through \nlo. They lead to the same set of master integrals as given in
\eqn{eq:masters}. Comparing our results to \citere{Luscher:2010iy} and
\citere{Makino:2014taa}\footnote{We compare to \texttt{arXiv} versions 2
  and 5 of that paper.}, we find full agreement.

\section{Coefficient functions through NNLO QCD}\label{sec:coefs}

The strong coupling and the quark mass require the regular
\qcd\ renormalization according to
\begin{equation}
  \begin{split}
    g_0 = \left( \frac{\mu \, e^{\EulerGamma / 2}}{\sqrt{4 \pi}}
    \right)^\ep Z_g\,g\,,\qquad m_0 = Z_m\,m\,,
    \label{eq:msbar}
  \end{split}
\end{equation}
where we write the renormalization constants $Z_g$ and $Z_m$ as
\begin{equation}
  \begin{split}
    Z_g &=
    1 - \gcoup{2}\frac{\beta_0}{2\ep} +
    \gcoup{4}\left(\frac{3\beta_0^2}{8\ep^2}
    -\frac{\beta_1}{4\ep}\right)   + \order{g^6}\,,
    \\
    Z_m &=
    1
    -\gcoup{2}\frac{\gamma_{m,0}}{2\ep}
    +\gcoup{4}\left[\frac{1}{\ep^2}\left(
      \frac{\gamma_{m,0}^2}{8} + \frac{\beta_0\gamma_{m,0}}{4}\right)
      -\frac{\gamma_{m,1}}{4\ep}\right] + \order{g^6}\,,
    \label{eq:rgconsts}
  \end{split}
\end{equation}
with
\begin{equation}
  \begin{split}
      \beta_0 &= \frac{11}{3} C_A - \frac{4}{3} \TF\,, \qquad
      \beta_1 = \frac{34}{3} C_A^2 - \left( 4 C_F + \frac{20}{3} C_A
      \right) \TF \,,\\
      \gamma_{m,0} &= 6 C_F\,,\qquad
      \gamma_{m,1} = \frac{97}{3} C_A C_F + 3 C_F^2
      - \frac{20}{3} C_F \TF \,.
    \label{eq:anombg}
  \end{split}
\end{equation}
$C_F$ and $C_A$ are the quadratic Casimir eigenvalues of the fundamental
and the adjoint representation of the gauge group,
respectively. Furthermore, $\TF=Tn_F$, with $n_F$ the number of quark
flavors, and $T$ the trace normalization in the fundamental
representation. For SU($N_c$), it is $C_F=(N_c^2-1)/(2N_c)$, $C_A=N_c$,
and $T=1/2$.

In addition, the flowed quark fields also require renormalization
according to
\begin{equation}
  \begin{split}
    \chi_{f,R} = \sqrt{Z_\chi}\,\chi_{f}\,,
  \end{split}
\end{equation}
leading to the factor $Z_\chi$ in the definition of the operators
$\tcalo_{3,4}$ in \eqn{eq:tops}.  $Z_\chi$ differs from the quark-field
renormalization of regular \qcd.
Through \nlo, the \msbar\ result has been evaluated in
\citere{Luscher:2013cpa}. To determine the renormalization constant at
\nnlo\ as required by our calculation, we may use the fact that the
coefficient $c_3(t)$ must be finite after renormalization. Writing the
\msbar\ expression as
\begin{equation}
  \begin{split}
    Z^{-1}_\chi &=
    1
      -\gcoup{2}\frac{\gamma_{\chi,0}}{2\ep}
      +\gcoup{4}\left[\frac{1}{\ep^2}\left(
        \frac{\gamma_{\chi,0}^2}{8} +
        \frac{\beta_0\gamma_{\chi,0}}{4}\right)
        -\frac{\gamma_{\chi,1}}{4\ep}\right] + \order{g^6}\,,
    \label{eq:zchi}
  \end{split}
\end{equation}
we find\footnote{Note that since $c_4=0$ at leading
  order, this coefficient only requires \nlo\ renormalization.}
\begin{equation}
  \begin{split}
  \gamma_{\chi,0} &= 6C_F\,,\\
  \gamma_{\chi, 1} &=
  C_A C_F \left( \frac{223}{3} - 16 \ln 2 \right) - C_F^2 \left( 3 + 16
  \ln 2 \right) - \frac{44}{3} C_F \TF\,.
  \end{split}
\end{equation}
This allows us to evaluate the coefficients of the energy-momentum
tensor in the \msbar\ scheme through \nnlo\ \qcd:\footnote{For
  convenience of the reader, we provide the expressions for
  $c_1,\ldots,c_4$ also in electronic form in an ancillary file with
  this paper.}
\begin{equation}
  \begin{split}
    c_1(t) =& \, \frac{1}{g^2} \Bigg\{ 1 + \gcoup{2}
    \left[ - \frac{7}{3}  C_A + \frac{3}{2}
      \TF -  \beta _0 \, L (\mu, t) \right] \\
	& \quad + \gcoup{4} \Bigg[ - \beta_1 \, L
      (\mu, t) + C_A^2 \left( - \frac{14482}{405} - \frac{16546}{135}
      \ln 2 + \frac{1187}{10} \ln 3 \right) \\
      & \qquad + C_A \TF \Bigg( \frac{59}{9}
      \text{Li}_2\left(\frac{1}{4}\right) + \frac{10873}{810} +
      \frac{73}{54} \pi^2 - \frac{2773}{135} \ln 2 + \frac{302}{45} \ln
      3 \Bigg) \\
	& \qquad + C_F \TF \Bigg( - \frac{256}{9}
      \text{Li}_2\left(\frac{1}{4}\right) +\frac{2587}{108} -
      \frac{7}{9} \pi^2 - \frac{106}{9} \ln 2 - \frac{161}{18} \ln 3
      \Bigg) \Bigg] \\
    & \quad + \order{g^6} \Bigg\} \, ,
    \label{eq:c1}
  \end{split}
\end{equation}
\begin{equation}
	\begin{split}
          c_2(t) =& \, \frac{1}{4 g^2} \Bigg\{ - 1 +  \gcoup{2}
          \left[ \frac{25}{6}  C_A - 3  \TF + \beta _0 \,
      L (\mu, t) \right] \\
	& \quad +  \gcoup{4} \Bigg[ \beta _1 \, L (\mu, t) +
      C_A^2 \left(\frac{56713}{1620} - \frac{1187}{10} \ln 3 +
      \frac{16546}{135} \ln 2 \right) \\
	& \qquad + C_A  \TF \Bigg( - \frac{59}{9}
      \text{Li}_2\left(\frac{1}{4}\right) - \frac{6071}{405} -
      \frac{73}{54} \pi^2 + \frac{2287}{135} \ln 2 - \frac{361}{90} \ln
      3 \Bigg) \\
	& \qquad + C_F \TF \Bigg( \frac{220}{9}
      \text{Li}_2\left(\frac{1}{4}\right) - \frac{1757}{54} +
      \frac{10}{9} \pi ^2 - \frac{164}{9} \ln 2 + \frac{247}{9} \ln 3
      \Bigg) \Bigg] \\
	& \quad  + \order{g^6} \Bigg\} \, ,
    \label{eq:c2}
  \end{split}
\end{equation}
\begin{equation}
	\begin{split}
	  c_3 (t) =& \, \frac{1}{4} \Bigg\{1 + \gcoup{2}
          \left( \frac{3}{2} C_F +
          \frac{\gamma_{\chi, 0}}{2} L (\mu, t) \right) \\
			& \quad + \gcoup{4}
          \Bigg[ \frac{\gamma_{\chi, 0}}{4} \left(\beta _0+
            \frac{\gamma_{\chi, 0}}{2} \right) \Big( L^2 (\mu, t) + L
            (\mu, t) \Big) + \frac{\gamma_{\chi, 1}}{2} L (\mu, t) \\
			& \qquad + C_F^2 \Bigg( - \frac{137}{9}
            \text{Li}_2\left(\frac{1}{4}\right) - \frac{559}{216} +
            \frac{103}{108} \pi^2 - \frac{1736}{27} \ln 2 +
            \frac{122}{3} \ln 3 - 4 \ln^2 2 \Bigg) \\
			&  \qquad + C_F  \TF \Bigg( -
            \frac{136}{9}  \text{Li}_2\left(\frac{1}{4}\right) -
            \frac{3377}{810} - \frac{7}{9} \pi^2 + \frac{1232}{135} \ln
            2 - \frac{136}{15} \ln 3 \Bigg) \\
			& \qquad + C_A C_F \Bigg( - \frac{365}{9}
            \text{Li}_2 \left(\frac{1}{4}\right) + \frac{261829}{3240} +
            \frac{77}{108} \pi^2 + \frac{5788}{45} \ln 2 \\
			& \qquad \quad - \frac{2102}{15} \ln 3 - 4 \ln^2
            2 \Bigg) \Bigg] \\
			& \quad + \order{g^6} \Bigg\} \, ,
    \label{eq:c3}
	\end{split}
\end{equation}
\begin{equation}
  \begin{split}
    c_4 (t) =& \, \frac{C_F}{2} \, \Bigg\{ \gcoup{2} +
    \gcoup{4} \Bigg[ \left( \beta _0
      + \frac{\gamma_{\chi, 0}}{2} \right) L (\mu, t) \\
      & \qquad + C_F \Bigg( - \frac{161}{18}
      \text{Li}_2\left(\frac{1}{4}\right) - \frac{41}{54} -
      \frac{55}{108} \pi^2 - \frac{1105}{27} \ln 2 +
      \frac{101}{6} \ln 3 \Bigg) \\
      & \qquad + \TF \Bigg( \frac{25}{9}
      \text{Li}_2\left(\frac{1}{4}\right) -
      \frac{20573}{1620} + \frac{5}{18} \pi^2 +
      \frac{6559}{135} \ln 2 - \frac{679}{30} \ln 3 \Bigg)
      \\
      & \qquad + C_A \Bigg( \frac{257}{36}
      \text{Li}_2\left(\frac{1}{4}\right) - \frac{137}{405}
      + \frac{11}{216} \pi^2 - \frac{419}{90} \ln 2 +
      \frac{1157}{60} \ln 3 \Bigg) \Bigg] \\
    & \quad + \order{g^6} \Bigg\} \, ,
    \label{eq:c4}
  \end{split}
\end{equation}
where we introduced the parameter\footnote{This parameter is motivated
  by the product of the typical factor $(8\pi t)^{\ep}$ occurring in
  flow-time integrals\,\cite{Luscher:2010iy}, and the usual definition
  of the renormalization scale in the \msbar\ scheme, see
  \eqn{eq:msbar}: $(8\pi t)^\ep (\mu^2e^{\EulerGamma}/(4\pi))^\ep = 1 +
  \ep\,L(\mu,t) + \order{\ep^2}$.}
\begin{equation}
	L (\mu, t) \equiv \ln \left( 2 \mu^2 t \right) + \EulerGamma\,,
        \label{eq:lmut}
\end{equation}
with the Euler-Mascheroni constant $\EulerGamma=0.57721\ldots$. Through
\nlo, these results are in full agreement with those of
\citere{Suzuki:2013gza,Makino:2014taa}. We have carried out the
calculation in the general $R_\xi$ gauge of regular \qcd; the fact that
the gauge-parameter dependence cancels in the final result serves as
another welcome check. The gauge parameter $\kappa$ of \eqn{eq:flow} has
been set to 1.

While the energy-momentum tensor $T_{\mu\nu}(x)$ is
renormalization-scheme independent, this is not necessarily the case for
the operators $\tcalo_{i,\mu\nu}(t,x)$ and the coefficient functions
$c_i(t)$. Since $\tcalo_{1,\mu\nu}$ and $\tcalo_{2,\mu\nu}$ do not
require operator renormalization, their matrix elements as well as the
coefficient function are indeed renormalization-scheme independent.  On
the other hand, using the quark-field renormalization $Z_\chi$ of
\eqn{eq:zchi} in the \msbar\ scheme, matrix elements of
$\tcalo_{i,\mu\nu}$ and coefficient functions $c_i(t)$ become
explicitly dependent on the renormalization scale $\mu$ for
$i\in\{3,4\}$.

\begin{figure}
  \begin{center}
    \begin{tabular}{cc}
      \includegraphics[width=.45\textwidth,bb=20 0 300 220]{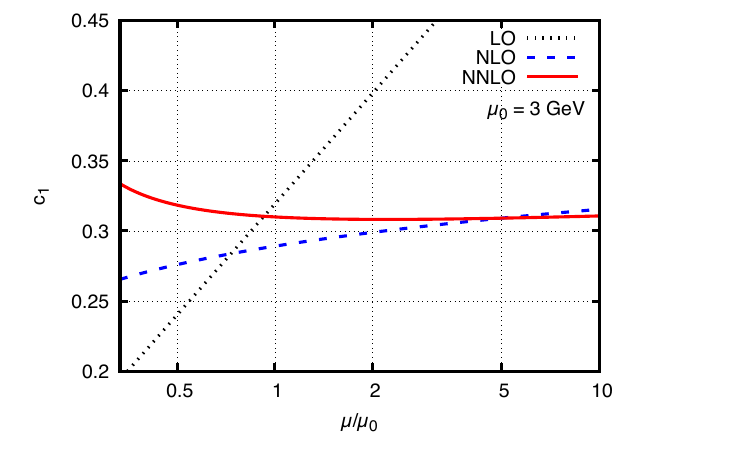} &
      \includegraphics[width=.45\textwidth,bb=20 0 300 220]{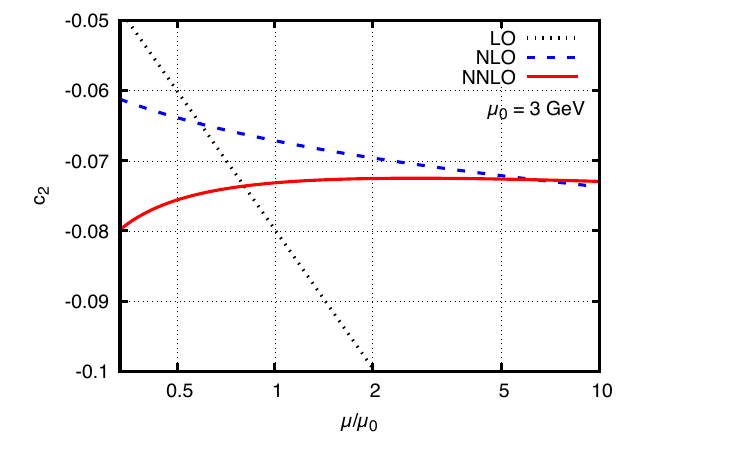} \\
      \includegraphics[width=.45\textwidth,bb=20 0 300 220]{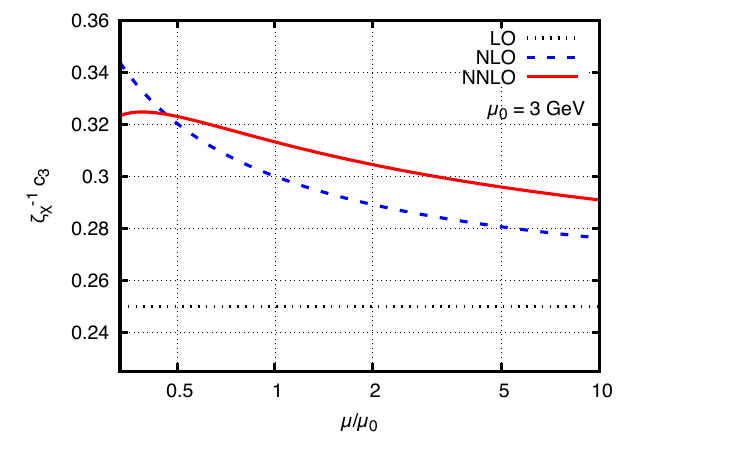} &
      \includegraphics[width=.45\textwidth,bb=20 0 300 220]{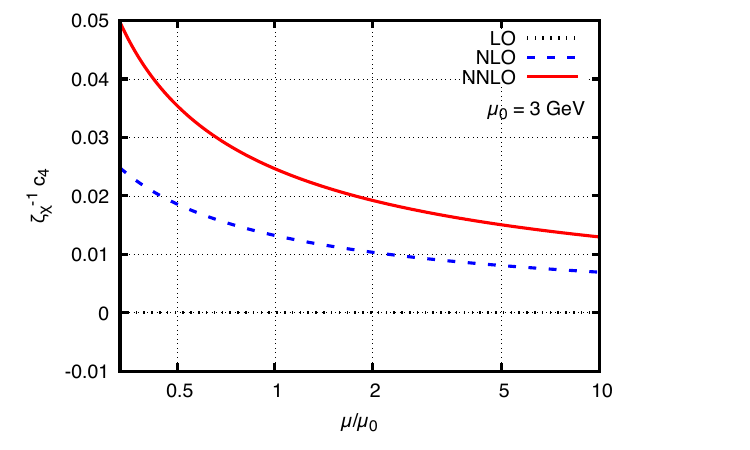}
    \end{tabular}
  \end{center}
	\caption{Renormalization-scale dependence of the coefficients
          $c_1$, $c_2$, $\ringc_3=\zeta^{-1}_\chi c_3$,
          $\ringc_4=\zeta^{-1}_\chi c_4$, defined in
          \eqs{eq:c1}--\noeqn{eq:c4}, with $\zeta_\chi$ from
          \eqn{eq:zfchi}. The dotted black, dashed blue, and solid red
          curve correspond to keeping terms up to order $(g^2)^{n-1}$ in
          $c_1$ and $c_2$, and $(g^2)^{n}$ in $\ringc_3$ and
          $\ringc_4$, with $n=0,1,2$, respectively. The central
          scale is set to $\mu_0=3$\,GeV, corresponding to $t = 3.12
          \cdot 10^{-2} /\text{GeV}^2$, see \eqn{eq:mu0}.
          The number of flavors is set to $n_F=3$.}
	\label{fig:mu3gev}
\end{figure}

\begin{figure}
\begin{center}
    \begin{tabular}{cc}
      \includegraphics[width=.45\textwidth,bb=20 0 300 220]{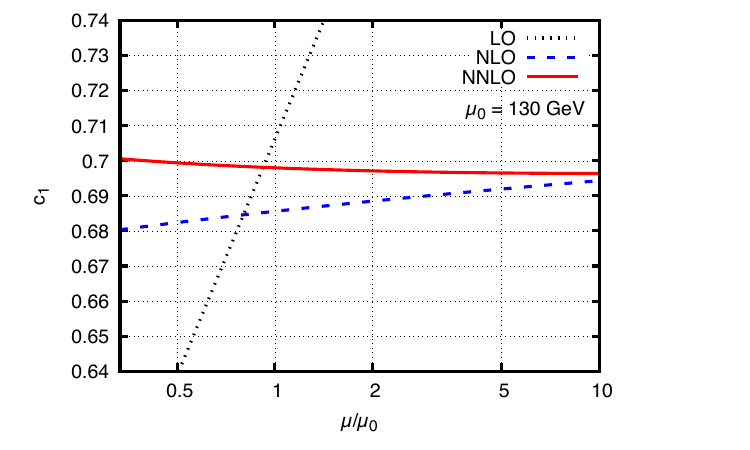} &
      \includegraphics[width=.45\textwidth,bb=20 0 300 220]{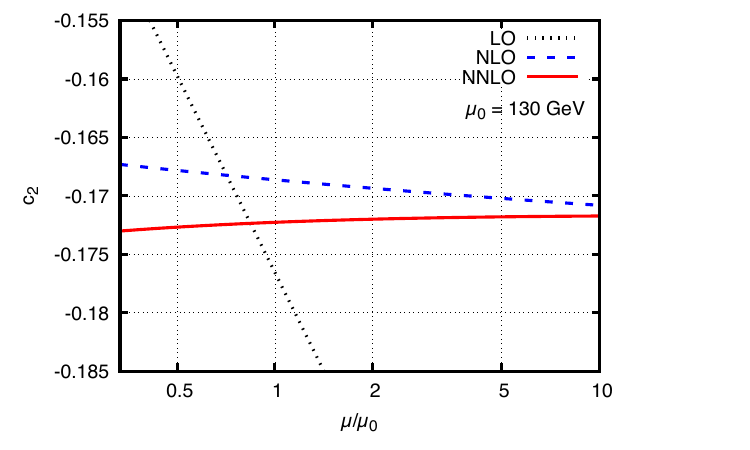} \\
      \includegraphics[width=.45\textwidth,bb=20 0 300 220]{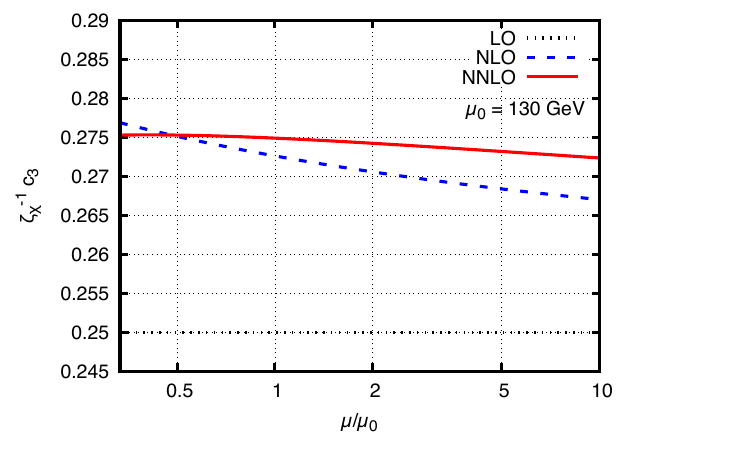} &
      \includegraphics[width=.45\textwidth,bb=20 0 300 220]{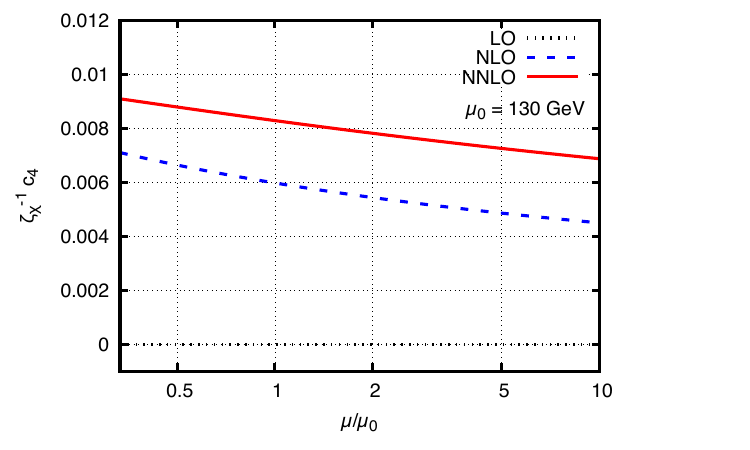}
    \end{tabular}
  \end{center}
	\caption{Same as \fig{fig:mu3gev}, but for $\mu_0=130$\,GeV (or
          $t=1.66\cdot 10^{-5} /\text{GeV}^2$), and $n_F=5$.}
	\label{fig:mu130gev}
\end{figure}

However, this renormalization-scheme dependence can be avoided by
introducing so-called ``ringed'' quark fields as suggested
in~\citere{Makino:2014taa}. This corresponds to replacing $Z_\chi$ in
\eqn{eq:tops} by
\begin{equation}
  \begin{split}
    \mathring{Z}_\chi(t) = \frac{-2 N_c \, n_F}{(4 \pi t)^2
      \langle\bar{\chi}_f(t,x)\overleftrightarrow
      {\slashed D}\chi_f(t,x)\rangle}\,.
    \label{eq:zf}
  \end{split}
\end{equation}
Currently, $\mathring{Z}_\chi(t)$ is available only through \nlo\ \qcd.
Its explicitly $\mu$-dependent terms can be reconstructed from the
requirement that $c_i(t)$ must be finite and $\mu$-independent for $i\in
\{3,4\}$ though. In this way we find for the ratio to the
\msbar\ quark-field renormalization constant of \eqn{eq:zchi}:
\begin{equation}
	\begin{split}
	  \zeta_\chi&\equiv \frac{\mathring{Z}_\chi}{Z_\chi} = 1 + \gcoup{2}
          \left( \frac{\gamma_{\chi, 0}}{2} L(\mu ,t) - 3 C_F \ln 3 - 4
          C_F \ln 2 \right) \\ &
                \qquad + \gcoup{4} \Bigg\{
                  \frac{\gamma_{\chi, 0}}{4} \left( \beta_0 +
                  \frac{\gamma_{\chi, 0}}{2} \right) L^2 (\mu, t) +
                  \Big[ \frac{\gamma_{\chi, 1}}{2} -
                  \frac{\gamma_{\chi, 0}}{2} \left( \beta_0 +
                  \frac{\gamma_{\chi, 0}}{2} \right) \ln 3 \\
                & \qquad \qquad -
                  \frac{2}{3}
                  \gamma_{\chi, 0} \left(
                  \beta_0 + \frac{\gamma_{\chi, 0}}{2}\right) \ln 2
                  \Big] L(\mu,t)
                  + C_2 \Bigg\} \\ &\qquad + \order{g^6} \,.
                \label{eq:zfchi}
	\end{split}
\end{equation}
The constant $C_2$ cannot be determined in this way, but requires a
dedicated \three-loop calculation of the \two-point function occurring
in the denominator of \eqn{eq:zf}. A detailed outline of this
calculation is beyond the scope of this paper; it will be presented
together with a more complete description of our setup in a forthcoming
publication\,\cite{Artz:2019bpr}. At this point, we simply quote the
numerical value of this result up to three significant digits, which is
more than sufficient in the light of the theoretical uncertainties to be
discussed below:\footnote{Note that the calculation of
  $\mathring{Z}_\chi$ also provided an independent check for $Z_\chi$.}
\begin{equation}
  C_2 = -23.8 \, C_A C_F + 30.4 \, C_F^2 - 3.92 \, C_F T_F\,.
\end{equation}
Multiplication of $c_3(t)$ and
$c_4(t)$ in Eqs.\,(\ref{eq:c3}) and (\ref{eq:c4}) by this ratio makes
also these coefficients formally $\mu$-independent, i.e.,
\begin{equation}
  \begin{split}
    \mu\dderiv{}{\mu}\{c_1,c_2,\ringc_3,\ringc_4\}=0\,,\quad
    \text{where}\quad \ringc_i\equiv \zeta_\chi^{-1} c_i\,.
  \end{split}
\end{equation}
As in any perturbative calculation, the $\mu$-independence only holds up
to higher orders in $g$. The decrease of the residual $\mu$-dependence
is thus commonly used as a qualitative check of the perturbation
expansion for the specific observable under consideration. We thus study
the $\mu$-dependence of the four coefficients after dividing $c_3$
and $c_4$ by the ratio $\zeta_\chi$ defined in \eqn{eq:zfchi}. We fix a
characteristic value for the flow time $t$ and vary the renormalization
scale $\mu$ around the central value $\mu_0$, which we define such that
$L(\mu_0,t)=0$, cf.\ \eqn{eq:lmut}, i.e.
\begin{equation}
  \begin{split}
    \mu_0 = \frac{e^{-\EulerGamma/2}}{\sqrt{2t}}\,.
    \label{eq:mu0}
  \end{split}
\end{equation}
Figures\,\ref{fig:mu3gev} and \ref{fig:mu130gev} show the leading order
(\lo), \nlo, and the \nnlo\ approximation of $c_1$, $c_2$, $\ringc_3$,
and $\ringc_4$ as functions of the renormalization scale for two
different values of the flow time $t$, corresponding to $\mu_0=3$\,GeV
and $\mu_0=130$\,GeV, respectively. In the former case, we set $n_F=3$,
in the latter $n_F=5$. We use $\alpha^{(n_F=5)}_s(M_Z)=0.118$ in order
to evaluate the input values for the couplings,
$g^{(n_F=3)}(3\,\text{GeV})=1.77$ and
$g^{(n_F=5)}(130\,\text{GeV})=1.19$.  The $\mu$-variation of the strong
coupling constant $g(\mu)$ is determined by numerically solving the
corresponding renormalization-group equation with the help of
\texttt{RunDec}\,\cite{Chetyrkin:2000yt,Herren:2017osy} at \one-, \two-,
and \three-loop level for the \lo, \nlo, and the \nnlo\ curve,
respectively.  In \fig{fig:mu3gev}, the value of $t$ is chosen such that
the central scale of \eqn{eq:mu0} is $\mu_0=3$\,GeV. At this central
scale, the \nnlo\ corrections increase the modulus of the coefficients $c_1$ and $c_2$ by 10\% and 13\%
relative to NLO, respectively. This is within twice the
\nlo\ uncertainty due to missing higher-order effects as estimated by
varying $\mu/\mu_0$ between 1/2 and 2, where one finds $7.3$\% for
$c_1$, and $8.0\%$ for $c_2$. We are therefore confident that the
\nnlo\ uncertainty estimated in the same way is rather reliable: it is
given by $5.7$\% for $c_1$ and $7.2$\% for $c_2$. Note that the dominant
contribution to these numbers comes from the downward variation of
$\mu$, where $g(\mu)$ starts to become sensitive to the non-perturbative
region. The behavior of $c_1$ and $c_2$ towards larger values of $\mu$
seems to suggest that this uncertainty estimate may actually be too
conservative.

As opposed to the gluonic coefficients $c_1$ and $c_2$, the coefficients
of the fermionic operators $c_3$ and $c_4$ exhibit a residual scale
dependence only from the \nlo\ term onwards. One therefore expects a
stronger $\mu$-dependence at \nnlo\ for these terms. Nevertheless, for
$\ringc_3$, the estimate of the theory uncertainty due to scale
variation still decreases from $9.8$\% to $8.1$\%. The increase of the
result due to the \nnlo\ effects is $8.3$\% relative to the \nlo\ result
at $\mu=\mu_0$.

The behavior of $\ringc_4$, on the other hand, is less
satisfactory at $\mu_0=3$\,GeV. The \nnlo\ effects more than double the
\nlo\ result in this case, and the uncertainty estimate due to scale
variation actually \textit{increases} from 47\% to 71\% when going from
\nlo\ to \nnlo.  Note, however, that $c_4=0$ at \lo, which means that
this coefficient is numerically sub-dominant.

As one would expect, for $\mu_0=130$\,GeV, the perturbative behavior of
all coefficients is significantly improved, cf.\ \fig{fig:mu130gev}. For
$c_1$, $c_2$, and $\ringc_3$, the scale uncertainty is at the
sub-percent level already at \nlo; at \nnlo, it amounts to less than
$0.2$\% in all three cases. The effect of the \nnlo\ corrections
relative to the \nlo\ result is about $2$\% for $c_1$ and $c_2$, and
$0.8$\% for $\ringc_3$. Also for $\ringc_4$, the
situation improves significantly: the \nnlo\ terms add 38\% to the
\nlo\ result, and the uncertainty goes down from 10\% at \nlo\ to
$5.8$\% at \nnlo.

It is also worth pointing out that the choice of the central scale
$\mu_0$ as defined in \eqn{eq:mu0} seems justified by the behavior of
the successively higher orders. In almost all cases, the \nlo\ and the
\nnlo\ corrections are \textit{both} relatively small at $\mu=\mu_0$. At
the same time, the \nnlo\ corrections relative to the \nlo\ result are
always smaller than the \nlo\ corrections compared to the
\lo\ result. The only exception to this is $\ringc_4$ at $\mu_0=3$\,GeV,
where, however, no choice of $\mu$ seems to stand out over any other.

In summary, we conclude that the \nnlo\ terms lead to a significant
improvement of the perturbative accuracy of the Wilson coefficients.

\section{Trace anomaly}\label{sec:traceanom}

As a test of our result, we use the trace anomaly of the \emt. As
suggested in \citere{Nielsen:1977sy}, a simple derivation consists of
taking the trace of the \emt\ in $D=4 - 2 \ep$ dimensions. By use of the
equations of motion, this gives for the gauge invariant part
\begin{equation}
	T_{\mu \mu} = \frac{\ep}{2 g_0^2} F_{\rho \sigma}^a F_{\rho
          \sigma}^a - \sum_{f=1}^{n_F} m_{f, 0} \overline{\psi}_f \psi_f =
        \frac{1}{2D}\left(\frac{\ep}{g_0^2}
        \calo_{2,\mu\mu}+\calo_{4,\mu\mu}\right)
        \,,
	\label{eq:EMTabare}
\end{equation}
where we have used \eqn{eq:eom} in the last step.
Using
the mixing matrix $\zeta_{i j}(t)$, we can rewrite this in terms of
flowed operators:
\begin{equation}
\begin{split}
  T_{\mu\mu} =  \bar c_i(t)\tcalo_{i,\mu\mu}(t,x)\,,\qquad
  \bar c_i(t) = \frac{1}{2D}\left(
  \frac{\ep}{ g_0^2} \zeta_{2 i}^{-1}(t)
         + \zeta_{4 i}^{-1}(t)\right)\,.
	 \label{eq:EMTaflowed}
  \end{split}
\end{equation}
Note that $\bar{c}_1 (t) = \bar{c}_3 (t) = 0$, as $\tcalo_{1, \mu \nu} (t, x)$
and $\tcalo_{3, \mu \nu} (t, x)$ have a non-trivial index structure and
therefore $\calo_{2, \mu \nu} (x)$ and $\calo_{4, \mu \nu} (x)$ cannot
mix with them.
Since $\tcalo_{2,\mu\mu}=D\tcalo_{1,\mu\mu}$ and
$2\tcalo_{4,\mu\mu}=D\tcalo_{3,\mu\mu}$, we cannot equate coefficients
with \eqn{eq:emtflow} for all $i$ individually. Instead, only the weaker
conditions
\begin{equation}
  \begin{split}
    c_1(t)+Dc_2(t) = D\bar c_2(t)\,,\qquad
    2c_3(t)+Dc_4(t) = D\bar c_4(t)\,.
  \end{split}
\end{equation}
can be derived. We checked that these equations are indeed fulfilled by
our result.

\section{Operator renormalization}\label{sec:zij}

Using the fact that flowed operators are finite after mass and field
renormalization, we can also compute the renormalization matrix for the
regular \qcd\ operators $\{\calo_{1,\mu\nu}(x),$ $\calo_{2,\mu\nu}(x),
\calo_{3,\mu\nu}(x), \calo_{4,\mu\nu}(x) \}$ defined in \eqn{eq:ops}. It
is convenient to define an equivalent set of operators as
\begin{equation}
  \begin{split}
    &\hat\calo_{i,\mu\nu}(x) = H_{ij}\calo_{j,\mu\nu} (x)\,,\qquad
    \mbox{where}\quad
    H_{ij} = \left\{\begin{array}{l}
    1/g_0^2\quad \text{for}\ i=j\in\{1,2\}\,,\\
    1\quad \text{for}\ i=j\in\{3,4\}\,,\\
    0\quad \text{for}\ i\neq j
    \end{array}\right.\,.
    \label{eq:opshat}
  \end{split}
\end{equation}
This multiplication of $\calo_{1,\mu\nu}$ and $\calo_{2,\mu\nu}$ by
$1/g_0^2$ ensures that the mass dimension of all operators
$\hat\calo_{i,\mu\nu}$ is equal to $D$.  The renormalization matrix is
then defined as
\begin{equation}
	\{ \hat\calo_i \}_R (x) = Z_{ij} \hat\calo_j (x) \, .
\end{equation}
Expressing the $\hat\calo_{i,\mu\nu} (x)$ in terms of flowed operators,
one can determine its entries in the \msbar\ scheme by demanding that
\begin{equation}
	 \{ \hat\calo_i \}_R (x) = Z_{i j} \, H_{j k} \,
         \zeta_{k l}^{-1} (t) \, \tcalo_l (t, x)
\end{equation}
be finite.  In analogy to \eqs{eq:rgconsts} and \noeqn{eq:zchi}, we write
\begin{equation}
	Z_{i j} = \delta_{i j} - \gcoup{2} \frac{\gamma_{0,
            i j}}{2 \ep} + \gcoup{4} \left[ \frac{1}{\ep^2}
          \left( \frac{\gamma_{0, i k} \gamma_{0, k j}}{8} +
          \frac{\beta_0}{4} \gamma_{0, i j} \right) - \frac{\gamma_{1, i
              j}}{4 \ep} \right] + \order{g^6}\,.
\end{equation}
Our result for the coefficients of the anomalous dimension at \nlo\ is
in agreement with \citere{Makino:2014taa}:
\begin{equation}
	\gamma_0 =
	\begin{pmatrix}
		- \frac{8}{3} \TF & \frac{11}{6} C_A &
                \frac{4}{3} C_F & \frac{7}{3} C_F \\ 0 & \frac{22}{3}
                C_A - \frac{8}{3} \TF & 0 & 12 C_F
                \\ \frac{32}{3}  \TF & - \frac{8}{3}  \TF &
                - \frac{16}{3} C_F & \frac{8}{3} C_F \\ 0 & 0 & 0 & 0
	\end{pmatrix}\,.
\end{equation}
At \nnlo, we find
\begin{equation}
\begin{aligned}
		\gamma_{1, 1 1} &= - \frac{140}{27} C_A \TF -
                \frac{296}{27} C_F \TF\,,  & \gamma_{1, 1 2}
                &=  \frac{34}{3} C_A^2 - \frac{112}{27} C_A
                \TF - \frac{10}{27} C_F  \TF\,, \\
		\gamma_{1, 1 3} &=  \frac{188}{27} C_A C_F -
                \frac{56}{27} C_F^2 - \frac{104}{27} C_F \TF\,, &
                \gamma_{1, 1 4} &=  \frac{812}{27} C_A C_F +
                \frac{85}{27} C_F^2 - \frac{44}{27} C_F \TF\,, \\
		\gamma_{1, 2 1} &= 0\,, & \gamma_{1, 2 2} &=
                \frac{136}{3} C_A^2 - \frac{80}{3} C_A  \TF - 16 C_F
                \TF \,, \\
		\gamma_{1, 2 3} &= 0\,, & \gamma_{1, 2 4} &=
                \frac{388}{3} C_A  C_F + 12 C_F^2 - \frac{80}{3} C_F
                \TF\,, \\
		\gamma_{1, 3 1} &= \frac{560}{27} C_A  \TF +
                \frac{1184}{27} C_F   \TF\,, & \gamma_{1, 3 2} &=
                - \frac{272}{27} C_A  \TF - \frac{392}{27} C_F
                \TF\,, \\
		\gamma_{1, 3 3} &= - \frac{752}{27} C_A  C_F +
                \frac{224}{27} C_F^2 + \frac{416}{27} C_F   \TF\,, &
                \gamma_{1, 3 4} &= \frac{244}{27} C_A C_F -
                \frac{16}{27} C_F^2 - \frac{544}{27} C_F   \TF\,, \\
		\gamma_{1, 4 1} &= \gamma_{1, 4 2} = \gamma_{1, 4 3} =
                \gamma_{1, 4 4} = 0\,.
    \label{eq:}
\end{aligned}
\end{equation}
The renormalization matrix $Z_{ij}$ and the mixing matrix $\zeta_{ij}$
are provided in ancillary files with this paper. In this way, one
obtains the following expression for the energy-density operator in
terms of flowed operators, for example:
\begin{equation}
  \begin{split}
    \bigg\{\frac{1}{g_0^2}&F_{\mu\nu}(x)F_{\mu\nu}(x)\bigg\}_R =
    \frac{Z_{2j}H_{jk}}{D}\left[\zeta_{k2}^{-1}(t)\,\tcalo_{2,\mu\mu}(t,x) +
      \zeta_{k4}^{-1}(t)\tcalo_{4,\mu\mu}(t,x)\right]\\
    &=\frac{1}{g^2} G_{\mu\nu}(t,x)G_{\mu\nu}(t,x) \Bigg\{ 1 -
         \frac{7}{2} C_A \gcoup{2} \\
      & \qquad + \gcoup{4}\Bigg[ \left( - \frac{3}{2}\, C_A^2 - 2\, C_A T_F - 14\, C_F T_F \right) L(\mu, t) \\
      & \qquad \quad + C_A^2 \left( - \frac{1427}{180} + \frac{87}{5} \ln 2 - \frac{54}{5} \ln 3 \right) + \frac{8}{9}\, C_A T_F - \frac{34}{3}\, C_F T_F \Bigg] \Bigg\} \\
    & \quad+ Z_\chi \sum_{f=1}^{n_F}\bar\chi_f(t,x) \overleftrightarrow{\slashed{\mathcal{D}}}\chi_{f}(t,x) \Bigg\{ \gcoup{2}C_F\left(5 + 6 L (\mu, t)\right) \\
    & \qquad + \gcoup{4} \Bigg[ \gamma_{\chi, 0} \left( \beta_0 + \frac{\gamma_{\chi, 0}}{2} \right) L^2(\mu, t) \\
    & \qquad\quad + \left( \frac{304}{3}\, C_A C_F + 18\, C_F^2 - \frac{80}{3}\, C_F T_F \right) L (\mu, t) \\
    & \qquad \quad+ C_A C_F \left( - 2 \text{Li}_2 \left( \frac{1}{4} \right) + \frac{2923}{30} - \frac{4}{3} \pi^2 + \frac{8546}{15} \ln 2 - \frac{2139}{5} \ln 3 \right) \\
    & \qquad\quad + 6\, C_F^2 - C_F T_F \left( 30 + \frac{4}{3} \pi^2 \right) \Bigg]
    \Bigg\}\,.
  \end{split}
\end{equation}
Through \nlo, this result agrees with
\citeres{Makino:2014taa,Suzuki:2018vfs}.
Similar relations can be derived for all other operators of
\eqn{eq:ops}, of course.

\section{Conclusions}\label{sec:conclusions}

We have presented the universal Wilson coefficients for the
gradient-flow definition of the energy-momentum tensor through
\nnlo\ \qcd. The \nnlo\ corrections modify the three numerically
dominant coefficients $c_1$, $c_2$, $\ringc_3$ at the level of 10\%
(1-2\%) for a central scale of $\mu_0=3$\,GeV ($\mu_0=130$\,GeV), where
$\mu_0$ is related to the flow time $t$ according to \eqn{eq:mu0}. We
observe a reduction of the theoretical uncertainty relative to the
\nlo\ result as derived from varying the renormalization scale by a
factor of two around its central value.  The behavior of the fourth
coefficient $\ringc_4$ is less satisfactory, but its impact is expected
to be numerically suppressed.

Aside from this main outcome, new results presented in this paper
include the flowed quark-field renormalization constant to \nnlo\ in the
\msbar\ scheme, and the anomalous dimension matrix for the regular
\qcd\ operators which make up the \emt.

In conclusion, we hope that our results will help to improve the studies
of the \emt\ on the lattice. They are the first outcome of a systematic
setup for higher-order perturbative calculations within the
gradient-flow formalism\,\cite{Artz:2019bpr}, which should prove useful also
in other applications of this theoretical framework.

\paragraph{Acknowledgments.}
We are indebted to Johannes Artz and Mario Prausa for helping to establish the
setup within which this calculation was performed. We would also like to thank
Tobias Neumann for fruitful communication, and for providing his tools which
enabled us to obtain $\mathring{Z}_\chi$ before the publication of
\citere{Artz:2019bpr}. Further thanks go to Mauro Papinutto and Alexandru
Sturzu for pointing out typos and erroneous plots in an earlier version of the
manuscript. This work was supported by \textit{Deutsche Forschungsgemeinschaft
  (DFG)}, project
\href{http://gepris.dfg.de/gepris/projekt/386986591}{386986591}.

\appendix

\section{Feynman rules}
\label{sec:frules}

In this section, we present the Feynman rules for the operators defined
in \eqn{eq:ops} in regular \qcd. Only the terms which are relevant for
the construction of the projectors in \eqs{eq:p1} and \noeqn{eq:pj} are
listed explicitly. All momenta are understood to be outgoing.
\begin{align}
  \calo_{1,\mu\nu} &= \partial_\mu A_\rho^c\partial_\nu A_\rho^c +
  \cdots&\Rightarrow&&
  \begin{gathered}
	\includegraphics{dias/O1.pdf}
  \end{gathered} =&
  - p_{1,\mu} p_{2,\nu} \delta_{\alpha\beta}\delta^{ab} \\
  \calo_{2,\mu\nu} &= 2 \delta_{\mu \nu} \partial_\rho A_\sigma^c\partial_\rho A_\sigma^c +
  \cdots&\Rightarrow&&
  \begin{gathered}
	\includegraphics{dias/O1.pdf}
  \end{gathered} =&
  - 4 \delta_{\mu \nu} p_1 \cdot p_2 \delta_{\alpha\beta}\delta^{ab} \\
  \calo_{3 f ,\mu\nu} &= \bar{\psi}_f \gamma_\mu \partial_\nu \psi_f +
  \cdots&\Rightarrow&&
  \begin{gathered}
	\includegraphics{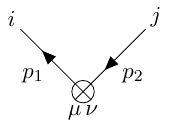}
  \end{gathered} =&
    \, i \gamma_\mu p_{2 \nu} \delta^{i j} \\
  \calo_{4 f,\mu\nu} &= \delta_{\mu \nu} \bar{\psi}_f \slashed{\partial} \psi_f +
  \cdots&\Rightarrow&&
  \begin{gathered}
	\includegraphics{dias/O3.pdf}
  \end{gathered} =&
    \, i \delta_{\mu \nu} \slashed{p}_2 \delta^{i j} \\
  \calo_{5 f,\mu\nu} &= \delta_{\mu \nu} m_0 \bar{\psi}_f \psi_f
  &\Rightarrow&&
  \begin{gathered}
	\includegraphics{dias/O3.pdf}
  \end{gathered} =&
  \, \delta_{\mu \nu} m_0 \delta^{i j}
    \label{eq:feynrules}
\end{align}

\def\app#1#2#3{{\it Act.~Phys.~Pol.~}\jref{\bf B #1}{#2}{#3}}
\def\apa#1#2#3{{\it Act.~Phys.~Austr.~}\jref{\bf#1}{#2}{#3}}
\def\annphys#1#2#3{{\it Ann.~Phys.~}\jref{\bf #1}{#2}{#3}}
\def\cmp#1#2#3{{\it Comm.~Math.~Phys.~}\jref{\bf #1}{#2}{#3}}
\def\cpc#1#2#3{{\it Comp.~Phys.~Commun.~}\jref{\bf #1}{#2}{#3}}
\def\epjc#1#2#3{{\it Eur.\ Phys.\ J.\ }\jref{\bf C #1}{#2}{#3}}
\def\fortp#1#2#3{{\it Fortschr.~Phys.~}\jref{\bf#1}{#2}{#3}}
\def\ijmpc#1#2#3{{\it Int.~J.~Mod.~Phys.~}\jref{\bf C #1}{#2}{#3}}
\def\ijmpa#1#2#3{{\it Int.~J.~Mod.~Phys.~}\jref{\bf A #1}{#2}{#3}}
\def\jcp#1#2#3{{\it J.~Comp.~Phys.~}\jref{\bf #1}{#2}{#3}}
\def\jetp#1#2#3{{\it JETP~Lett.~}\jref{\bf #1}{#2}{#3}}
\def\jphysg#1#2#3{{\small\it J.~Phys.~G~}\jref{\bf #1}{#2}{#3}}
\def\jhep#1#2#3{{\small\it JHEP~}\jref{\bf #1}{#2}{#3}}
\def\mpl#1#2#3{{\it Mod.~Phys.~Lett.~}\jref{\bf A #1}{#2}{#3}}
\def\nima#1#2#3{{\it Nucl.~Inst.~Meth.~}\jref{\bf A #1}{#2}{#3}}
\def\npb#1#2#3{{\it Nucl.~Phys.~}\jref{\bf B #1}{#2}{#3}}
\def\nca#1#2#3{{\it Nuovo~Cim.~}\jref{\bf #1A}{#2}{#3}}
\def\plb#1#2#3{{\it Phys.~Lett.~}\jref{\bf B #1}{#2}{#3}}
\def\prc#1#2#3{{\it Phys.~Reports }\jref{\bf #1}{#2}{#3}}
\def\prd#1#2#3{{\it Phys.~Rev.~}\jref{\bf D #1}{#2}{#3}}
\def\pR#1#2#3{{\it Phys.~Rev.~}\jref{\bf #1}{#2}{#3}}
\def\prl#1#2#3{{\it Phys.~Rev.~Lett.~}\jref{\bf #1}{#2}{#3}}
\def\pr#1#2#3{{\it Phys.~Reports }\jref{\bf #1}{#2}{#3}}
\def\ptp#1#2#3{{\it Prog.~Theor.~Phys.~}\jref{\bf #1}{#2}{#3}}
\def\ptep#1#2#3{{\it PTEP~}\jref{\textbf{#1}}{#2}{#3}}
\def\ppnp#1#2#3{{\it Prog.~Part.~Nucl.~Phys.~}\jref{\bf #1}{#2}{#3}}
\def\rmp#1#2#3{{\it Rev.~Mod.~Phys.~}\jref{\bf #1}{#2}{#3}}
\def\sovnp#1#2#3{{\it Sov.~J.~Nucl.~Phys.~}\jref{\bf #1}{#2}{#3}}
\def\sovus#1#2#3{{\it Sov.~Phys.~Usp.~}\jref{\bf #1}{#2}{#3}}
\def\tmf#1#2#3{{\it Teor.~Mat.~Fiz.~}\jref{\bf #1}{#2}{#3}}
\def\tmp#1#2#3{{\it Theor.~Math.~Phys.~}\jref{\bf #1}{#2}{#3}}
\def\yadfiz#1#2#3{{\it Yad.~Fiz.~}\jref{\bf #1}{#2}{#3}}
\def\zpc#1#2#3{{\it Z.~Phys.~}\jref{\bf C #1}{#2}{#3}}
\def\ibid#1#2#3{{ibid.~}\jref{\bf #1}{#2}{#3}}
\def\otherjournal#1#2#3#4{{\it #1}\jref{\bf #2}{#3}{#4}}
\newcommand{\jref}[3]{{\bf #1}, #3 (#2)}
\newcommand{\hepph}[1]{\href{http://arXiv.org/abs/hep-ph/#1}{\texttt{hep-ph/#1}}}
\newcommand{\hepth}[1]{\href{http://arXiv.org/abs/hep-th/#1}{\texttt{hep-th/#1}}}
\newcommand{\heplat}[1]{\href{http://arXiv.org/abs/hep-lat/#1}{\texttt{hep-lat/#1}}}
\newcommand{\mathph}[1]{\href{http://arXiv.org/abs/math-ph/#1}{\texttt{math-ph/#1}}}
\newcommand{\arxiv}[2]{\href{http://arXiv.org/abs/#1}{\texttt{arXiv:#1\,[#2]}}}
\newcommand{\bibentry}[4]{#1, {\it #2}, #3\ifthenelse{\equal{#4}{}}{}{, }#4.}

\IfFileExists{./\jobname_ref.tex}{
  
}{}

\end{document}